\documentclass[12pt,preprint]{aastex}
\newbox\grsign \setbox\grsign=\hbox{$>$} \newdimen\grdimen \grdimen=\ht\grsign
\newbox\laxbox \newbox\gaxbox
\setbox\gaxbox=\hbox{\raise.5ex\hbox{$>$}\llap
     {\lower.5ex\hbox{$\sim$}}}\ht1=\grdimen\dp1=0pt
\setbox\laxbox=\hbox{\raise.5ex\hbox{$<$}\llap
     {\lower.5ex\hbox{$\sim$}}}\ht2=\grdimen\dp2=0pt
\def\gax{\mathrel{\copy\gaxbox}}
\def\lax{\mathrel{\copy\laxbox}}
\def\as     {\ifmmode {\rlap.}$\,$''$\,$\! \else ${\rlap.}$\,$''$\,$\!$\fi}
\def\degg   {\ifmmode {\rlap.}$\,$^{\circ}$\,$\! \else ${\rlap.}$\,$^{\circ}$\,$\!$\fi}
\def\s      {\ifmmode {\rlap.}$\,$^{s}$\,$\! \else ${\rlap.}$\,$^{s}$\,$\!$\fi}
\def\kms    {\ifmmode{{\rm ~km~s}^{-1}}\else{~km~s$^{-1}$}\fi}
\def\eg     {e.g.,~}

\def\hho    {H$_2$O}

\def\apj    {{ApJ}}
\def\apjs   {{ApJS}}

\def\aap    {{A\&A}}

\shorttitle{New Class of OH Masers}
\shortauthors{Argon, Reid \& Menten}

\begin{document}
 
\title{A Class of Interstellar OH Masers Associated with Protostellar Outflows}
\author{Alice L. Argon and Mark J. Reid}
\affil{Harvard-Smithsonian Center for Astrophysics,
60 Garden Street/MS42, Cambridge MA 02138}

\author{Karl M. Menten}
\affil{Max-Planck-Institut f\"ur Radioastronomie,
Auf dem H\"ugel, D-53121 Bonn, Germany}

\begin{abstract}
        Using the Very Large Array, we have detected weak OH maser emission 
near the Turner--Welch protostellar source in the W3~OH region.  
Unlike typical interstellar OH masers, which are associated with
ultra-compact HII regions, our measured positions and 
proper motions (from Very Long Baseline Interferometry)
indicate that these OH masers are associated 
with a bipolar outflow traced by strong \hho\ masers.  These OH
masers may be part of a class of interstellar OH masers that are
associated with very young stars which have yet to, or may
never, create ultra-compact HII regions.
This class of OH masers appears to form near the edges of very dense 
material (within which \hho\ masers form), where 
total densities drop precipitously and interstellar UV radiation is 
sufficient to dissociate the \hho\ molecules.
Observations of this class of OH masers may be an important
way to probe the distribution of this important molecule
in interstellar shocks at arcsecond resolution or better.      
\end{abstract}

\keywords{ISM: individual (W3)---ISM: jets and outflows---masers---radiation
mechanisms: nonthermal---radio lines: ISM---stars: formation}

\section{Introduction}

        In the Milky Way, hydroxyl masers (OH) are found associated 
with either evolved stars (stellar masers), regions of massive star 
formation (interstellar masers), or the interface between SNRs and molecular 
material.  Interstellar OH masers are often observed 
from molecular material surrounding ultra-compact HII (UCHII) regions.
The OH masers in the source W3~OH are some of the strongest and best studied 
of their kind.  In this source, the OH maser spots
have been mapped with VLBI observations and found projected toward 
an UCHII region \citep{RHB80}.  Proper motions of these
masers \citep{BRM92} suggest a slow, general expansion 
with a velocity of a 
few km~s$^{-1}$.

        Interstellar \hho\ masers are often found in the same star 
forming regions as OH masers.  However, frequently the \hho\ masers
do not appear to be {\it directly} associated with UCHII regions.
In fact, rarely can one detect a stellar source at the center of the 
\hho\ maser flow, not even at IR wavelengths, suggesting a deeply embedded 
and probably very young object.  Moreover, \hho\ masers are also found 
in regions of low-mass star formation, which are devoid of OH masers. 
\hho\ masers often exhibit strong outflow motions, with expansion speeds
$\sim30$~km~s$^{-1}$, and sometimes high velocity features moving
at hundreds of km~s$^{-1}$.   

        In this paper we present a detailed analysis of weak OH masers  
in the W3~OH region that do {\it not} appear to be associated with an UCHII 
region.  Instead they seem to form at the edges of the 
outflow from the Turner--Welch object \citep{TW84}, 
a protostellar source or cluster of sources \citep{WSW99} 
that drives the \hho\ masers near W3~OH
\citep{AMM93} and a synchrotron jet \citep{RAM95,WRM99}.  Our observations 
suggest that protostellar outflows may be an important mechanism for the 
production of weak OH masers when no ionizing stars are present.

\section{Observations and Results}

\subsection{VLA Observations and Results}

W3~OH was observed in the 1612.231, 1665.4018, 1667.359, and 1720.530 MHz 
OH transitions with the NRAO\footnote{The National Radio Astronomy Observatory 
(NRAO) is operated by Associated Universities, Inc., under a cooperative 
agreement with the National Science Foundation.}  
Very Large Array (VLA) on 1992 December 27 using all 27
antennas in the A-configuration.  Three 1.5 minute scans were obtained in
the 1665 transition, three 2.5 minute scans in the 1667 transition,
and one 2.5 minute scan in each of the remaining two transitions.  When a
transition was observed more than once, scans were spread over about 3 hours.
Weather conditions were excellent.

Observations were made in both left-circular (LCP) and right-circular (RCP)
polarizations.  These observations were part of a project to map a large 
sample of interstellar OH masers \citep{ARM00}.  The 
observing bandwidth of 0.1953 MHz was centered at an 
LSR velocity of $-$44.0 km s$^{-1}$.  For each band the inner 128 spectral 
channels (out of a total of 256) were correlated, yielding a channel 
separation of 0.14 km s$^{-1}$.

Calibration and imaging were done in the Astronomical Image Processing System 
(AIPS).  Channel ``0'' observations (an average of the central 75$\%$ of 
each bandpass) were used to calibrate the complex instrumental 
gain for each antenna.  We employed the standard method of boot-strapping the 
flux density of a (variable) secondary calibrator, B0212$+$735, from
the flux density of the primary calibrator, 3C~286.  We then determined 
complex gains for the secondary calibrator using the AIPS task CALIB, 
interpolated those gains to the W3~OH observation times with the task CLCAL, 
and applied them to the full spectral-line database with the tasks TACOP and 
SPLIT.  Observations of the secondary calibrator were made immediately before 
and after each W3~OH scan to ensure accurate interpolation.

To remove the effects of any residual atmospheric phase corruption, we
``self-calibrated'' the data.  This was accomplished by selecting a strong
unresolved feature in one of the polarizations for each transition (called the
``reference feature''), shifting the interferometer phase center to the
position of that feature, determining residual phases using a point source, 
and subtracting those residual phases from the
data in all spectral channels in both polarizations.
The reference features chosen were the $v_{LSR}=-48.9$~\kms\ and 
$-44.4$~\kms\ channels at 1665 and 1667 MHz, respectively, both in 
LCP.  

A comparison of the positions of the strong OH masers in our study 
\citep{ARM00} and another study (\citet{BRM92} 
and references therein), taking into account the different map center 
locations, suggests that our absolute position error for the 1665 MHz 
transition is $\approx0\as1$ in each direction.  Self-calibration was followed 
by imaging, whereby each spectral channel was mapped and CLEANed.  
We searched an area of $128'' \times 128''$ ($512 \times 512$ pixels). 
Note that because our sources can be 100\% circularly polarized, we divided 
the maps produced by MX by a factor of two so that Stokes I is obtained by 
summing (rather than averaging) the RCP and LCP maps.

After imaging, selected peaks were fit with a two-dimensional Gaussian 
brightness distribution to determine their flux density, size, and 
offset from the reference feature.
While strong OH maser emission for the UCHII region has been mapped before,  
we searched for emission beyond the innermost $5'' \times 5''$.  
Spurious maser features were rejected by demanding that the
following four constraints be met:

\begin{itemize}
\item[] 1) {Peaks had to have a flux density of at least 4.5 times the channel 
rms noise.  Typical channel rms noise levels were 28 mJy for the 
1665.4018 MHz transition and 26 mJy for the 1667.359 MHz transition.}
\item[] 2) {Peaks had to have a flux density of at least 1.3 times the absolute 
value of the largest negative peak in the channel map.}
\item[] 3) {Peaks could not be any weaker than 1/150 of the peak in the map
owing to dynamic range limitations, which are particularly severe at 
1665 MHz because of the strong OH emission from the UCHII region
between $-43<v_{LSR}<-47$ ~km s$^{-1}$.  Two ``peaks'' appear in the spectrum 
shown in Figure 1 that do not satisfy this criterion.  They are artifacts of 
a 132 Jy ($-$46.3 km s$^{-1}$, LCP) and a 201 Jy ($-$45.0 km s$^{-1}$, RCP) 
feature.}
\item[] 4) {Peaks had to persist over three or more adjacent channels 
to within a synthesized beam width.}
\end{itemize}

\begin{figure}
\plotone{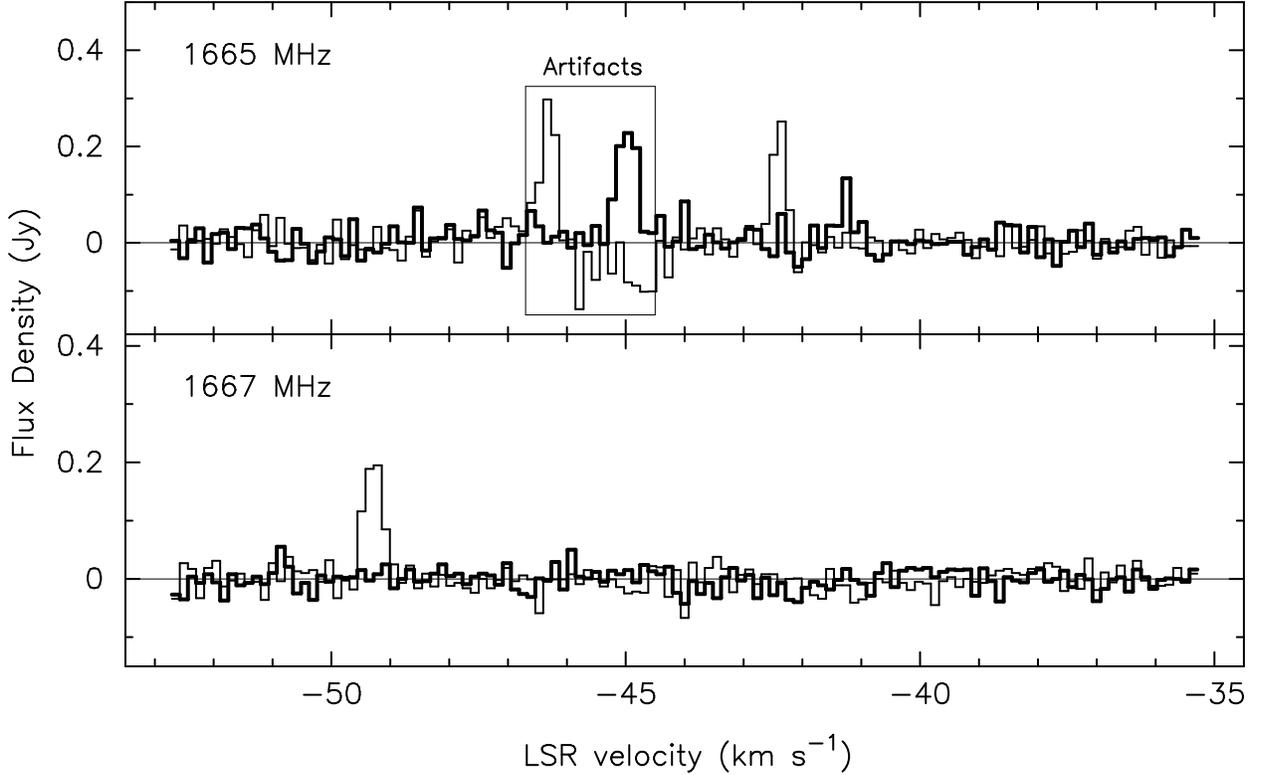}
\figcaption[]{
Spectra of two ``outlying'' OH maser features ($-42.4$~\kms, LCP at 1665 MHz 
and $-49.3$~\kms, LCP at 1667 MHz) at the positions of their maximum 
emission.  Heavy lines indicate RCP and light lines LCP emission.  The 
``features'' at v$_{LSR}$ $-46.3$~\kms\ (LCP) and $-45.0$~\kms\ (RCP) in the 
1665 MHz transition are artifacts generated by spatial sidelobes of the 
strongest 1665 LCP (132 Jy) and RCP (201 Jy) features, respectively, which are 
located toward the UCHII region.  The outlying OH features are {\it not} 
artifacts since there is no strong LCP emission at v$_{LSR}=-42.4$~\kms\ in 
1665 or at $-49.3$~\kms\ in 1667 elsewhere in the map.  A $4\sigma$ RCP 
detection at $-41.3$~\kms\ in the 1665 MHz transition may be part of a Zeeman 
pair, the other member being the $-42.4$~\kms\ feature. 
\label{fig1} 
}
\end{figure}

\begin{deluxetable}{lccccc}
\footnotesize
\tablecaption{VLA Observations of OH Maser Emission$^a$ \label{}}
\tablenum{1}
\tablewidth{0pt}
\tablehead{
\colhead{Transition (Pol.)}& \colhead{ $v_{\rm LSR}$$^b$ }& 
\colhead{Peak Flux Density$^b$}& \colhead{$\Delta v^b$}& 
\colhead{$\Delta\Theta_x^c$}& \colhead{$\Delta\Theta_y^c$}\nl
& \colhead{ (km s$^{-1}$) }& \colhead{(Jy)}& 
\colhead{ (km s$^{-1}$) }& \colhead{$('')$}& \colhead{$('')$}
}

\startdata

1665.4018            (LCP)& $-42.4$& $0.25\pm0.03$& $0.29$& $4.17\pm0.06$& 
$-0.09\pm0.06$\nl
1667.359\phantom{8}  (LCP)& $-49.3$& $0.20\pm0.03$& $0.37$& $6.81\pm0.07$& 
$-0.06\pm0.07$\nl

\tablecomments{
$^a$~~Maser features $>2\as5$ from center of the UCHII region in W3~OH.
\noindent
$^b$~~The quantities $v_{\rm LSR}$, $\Delta v$ (FWHM), and peak flux density 
were determined by fitting a Gaussian profile to the three highest spectral 
points.
\noindent  
$^c$~~Positions are offsets eastward ($\Delta\Theta_x$) and northward 
($\Delta\Theta_y$) from the map center of Fig.~2, located at 
$\alpha_{B1950} = 02^h 23^m 16\s499, \delta_{B1950} = 61^{\circ} 38' 
57\as21$, and are weighted averages of the contributing channels.  
Specifically, the position offsets are relative to the 
1665 MHz LCP reference feature at $-46.3$~\kms\ 
and then shifted by adding ($-0\as706,-0\as210$)  
to the relative maser positions to register them with the map in Fig.~2.
}

\enddata
\end{deluxetable}

\begin{deluxetable}{lccccc}
\footnotesize
\tablecaption{VLBI Observations of 1665 MHz LCP OH Maser Emission \label{}}
\tablenum{2}
\tablewidth{0pt}
\tablehead{
\colhead{Epoch}& \colhead{ $v_{\rm LSR}$ }& \colhead{Peak Flux Density}& 
\colhead{$\Delta v$}& \colhead{$\Delta\Theta_x$}& \colhead{$\Delta\Theta_y$}\nl
& \colhead{ (km s$^{-1}$) }& \colhead{(Jy)}& 
\colhead{ (km s$^{-1}$) }& \colhead{$('')$}& \colhead{$('')$}
}	

\startdata

1978& $-42.9$& $0.75\pm0.10$& $0.26$& $4.222\pm0.001$& $-0.052\pm0.001$\nl
1986& $-42.4$& $0.17\pm0.04$& $0.29$& $4.199\pm0.001$& $-0.094\pm0.001$\nl

\tablecomments{The positional errors are {\it formal} uncertainties.  See 
text for details and the Table 1 caption for other information.}

\enddata
\end{deluxetable}

We found two spots of OH maser emission satisfying all four criteria, one 
at 1665 and the other at 1667 MHz (see Table 1).  Since the expected 
uncertainty for a point source is approximately half the synthesized 
beam ($1''$) divided by the signal-to-noise ratio, the relative position 
error in each coordinate is $\approx0\as06$ and $\approx0\as07$ for the 
1665 and 1667 MHz transitions, respectively.  Spectra of the two features 
at the pixel of their maximum flux density are shown in Figure 1.  Both
features are $\gax8\sigma$ detections and were detected by Gaume \& Mutel
(1987), who, while noting that they were in the vicinity of the water masers
and HCN emission, did not discuss their significance.  Our $-42.4$ and $-49.3$ 
km s$^{-1}$ features were reported at $-42.8$ and $-48.3$ km s$^{-1}$, 
respectively, by Gaume \& Mutel, who list velocities of the peak channel 
emission in 1.1 km s$^{-1}$ wide channels.  Thus, the agreement in both 
position and velocity for these two measurements is reasonable.  Gaume \&  
Mutel also observed one additional spot of weak emission (0.1 Jy, a $3\sigma$ 
detection) {\it outside of} the central $5''$ x $5''$, that we fail to confirm.
Our $1\sigma$ noise level was 0.026 Jy in this transition, so if the maser's
flux density was constant (or increasing) in time between their observations
in 1985 and ours in 1992, we would have a $4\sigma$ detection.  Either Gaume
\& Mutel's detection is spurious, which is not highly unlikely in a
$\sim 10''$ x $10''$ field (containing about 100 independent beams), or the
feature weakened over the years.

\subsection{VLBI Data and Results}

In an attempt to confirm the 1665 MHz VLA detection and obtain a proper 
motion, we re-mapped the data from two epochs of VLBI observations: one 
in 1978 by \citet{GBR88} and the other in 1986 by \citet{BRM92}.  
(No archival VLBI data for the 1667.359 transition were available).  
At both epochs, RCP and LCP were observed and there 
were 96 spectral channels per IF, giving channel separations of 0.12 km 
s$^{-1}$.  Observations were centered at an LSR velocity of $-$46.0 km 
s$^{-1}$.  At both epochs, we searched for emission within
($\pm 0\as1$) of the positions in Table 1.  Since the features in W3~OH are weak
and heavily resolved on long baselines,
we used a (u--v)-taper of 20~M$\lambda$.  The resulting 
rms noise levels in the channel maps were about 100 mJy and 35 mJy for 1978 
and 1986 observations, respectively.  

The VLBI results are given in Table 2 and displayed in Figure 2.  The 1665 MHz 
LCP feature near 
$v_{LSR}=-42.4$~\kms\ in our VLA data was found to be present in the 1978 
and 1986 VLBI maps, albeit at a slightly different velocity ($-42.9$~\kms) in 
1978.  Note that \citet{GM87} report a velocity of $-42.8$~\kms\ for 
this feature.  We registered the maps at the three epochs (two VLBI, one VLA) 
by using the strongest 1665 MHz LCP feature (132 Jy at $-46.3$~\kms\ in the 
VLA data), which is located at $\alpha_{B1950} = 02^h 23^m 16\s322~\pm~0\s01, 
\delta_{B1950} = 61^{\circ} 38' 57\as66~\pm~0\as1$.  
The VLBI relative position errors 
in Table 2 are $\approx$~1 mas.  Assuming these features trace a single cloud 
of gas, we find a proper motion of about $-32$~\kms\ eastward and $-58$~\kms\ 
northward, with a {\it formal uncertainty} of about $2$~\kms\ in each 
direction.  A distance of 2.2 kpc was assumed.  Because the line center 
velocity of the feature changes by $\approx0.5$~\kms\ between the two epochs, 
a realistic proper motion uncertainty should include the possibility of 
structural changes in the feature.  The span of position offsets over the 
three or more adjacent channels that comprise a feature is one way to estimate 
the magnitude of possible structural changes.  A typical span was found to be 
$\approx2$~mas, which gives rise to a more {\it realistic proper motion 
uncertainty} of $\approx4$~\kms\ in each direction.  Only the two VLBI epochs 
were used to estimate the motion since their positional accuracies are far 
better than the positional accuracy of the VLA epoch.  The VLA position, 
however, is consistent with the derived VLBI motion and our OH motion is 
consistent with the bipolar outflow seen in the \hho\ masers \citep{AMM93}.  

Why does the Doppler velocity change over the eight years between our two
VLBI epochs?  The possibility that structural changes have occurred in the
feature was mentioned above.  A second possibility is that these masers, unlike
typical HII region masers, are likely to occur in a strongly accelerated
region, the region between the terminus of the H$_2$O outflow and the dense
ambient molecular cloud.  If this is the case, it would not be unreasonable 
to find a change of 0.5 km s$^{-1}$ in Doppler velocity between the two epochs.

Bloemhof, Reid \& Moran (1992) note a flux density calibration uncertainty of 
about 20$\%$ in the VLBI data.   Since this can not account for the 
difference among the three epochs, we conclude that 
this feature is variable.  

\section{Discussion}

\subsection {OH Masers Associated with the TW protostar}

The OH maser features described in this paper are not associated with the 
UCHII region in W3~OH, where strong OH and class II CH$_3$OH 
maser emission is found \citep{RHB80,BRM92,M91,MMW99}.  Instead,
as seen in Fig.~2 their positions, and the 1665 MHz feature proper motion, 
suggest an association with the TW object and its intense H$_2$O masers 
\citep{AMM93}.  The association of some OH masers in W3~OH with the TW object 
and its H$_2$O masers suggests that some interstellar OH 
masers that are offset from HII regions may be associated with protostellar 
outflows and very young stars which have yet to, or may never, ionize their 
surroundings.

The TW object is thought to be a newly formed or forming massive star 
\citep{TW84,WRM99} or stars \citep{WSW99},
which have yet to ionize their placental material.  
The ``hot core'' nature of the TW object was established by 
observations of high excitation NH$_3$~(5,5) inversion lines, which yield
an  NH$_3$ rotation temperature of 180~K \citep{MWW86}. 
While the TW object shows no detectable thermal bremsstrahlung, 
surprisingly, it is a source of synchrotron emission \citep{RAM95}. 
When imaged at high resolution ($0\as2$) and sensitivity (10~$\mu$Jy rms) 
with the VLA, the synchrotron source appears as an elongated, wiggling, 
double-sided jet \citep{WRM99}.  
The predominantly east--west orientation of the synchrotron jet 
matches both the spatial distribution and proper motions of the H$_2$O masers.
Also, the center of the synchrotron jet lies, within measurement 
uncertainties of $0\as1$, at the center of expansion of the
H$_2$O masers \citep{AMM93}.

Comparison of the positions and velocities of the TW source 
OH masers with the dust continuum and molecular line emission mapped by 
\citet{WHS97} and \citet{WSW99} also indicates a very close correspondence.
Both the 1665 and the 1667 MHz masers project on the edges of
the dust continuum/molecular emission.  The thermally excited molecular 
gas displays an apparent shift in velocity across
the source, with $v_{LSR}$ changing from about $-52$~km s$^{-1}$
at the eastern edge to about $-45$~km s$^{-1}$ at the western edge.
This trend of velocity increasing from east to west 
is consistent with our OH maser velocities.

The arrangement of the OH masers at, or just beyond, the ends of the 
H$_2$O maser outflow is noteworthy.   One possible explanation for this 
arrangement is that the H$_2$O molecules in the outflow are dissociated at the 
extremities of the source, producing OH molecules with sufficient density to 
yield detectable maser emission.  H$_2$O masers are estimated to require H$_2$ 
densities between $10^8$ and $10^{10}$ cm$^{-3}$, whereas OH masers require 
densities between $10^5$ and $10^{7}$ cm$^{-3}$ (\citet{E92} and references 
therein).  The upper limits to the H$_2$ densities are established by 
collisional thermalization of the level populations and quenching of the 
population inversions.  The lower limits arise from estimates of the minimum 
column density necessary to achieve strong maser amplification over reasonable 
path lengths.  Therefore, the transition from H$_2$O to OH masing in a source
like the TW object suggests a rapid decrease in density with
distance and a sharp boundary for the elongated region of H$_2$O maser
activity.  

Confirmation of a sharp molecular boundary comes from mm-wavelength
observations of dust continuum emission.
The molecular column density inside of the OH masers in the TW object, 
inferred from the dust continuum, of $N_{H_2} \sim 3\times 10^{24}$~cm$^{-2}$
indicates a volume density of $n_{H_2} \sim 10^8$~cm$^{-3}$, assuming
a line of sight path length of 0.01~pc (P. Schilke, pers. comm.).
The relatively sharp boundary of the dust continuum emission suggests
that the molecular density must drop by a factor of 10 or more
within a projected distance of about 0.005 pc at the edge of the source.  
Possibly, when molecular densities fall below $\sim10^7$~cm$^{-3}$, 
interstellar ultraviolet photons can dissociate H$_2$O molecules and 
create OH radicals.

\subsection {Comparison to Other Regions: Shock Chemistry}

While most interstellar OH masers do not show evidence of strong
outflows, there is conclusive evidence that \hho\
masers in regions of massive star formation partake in strong outflows 
from both 1) the large radial velocities observed, exceeding 
$\pm100$ \kms\ in Orion-KL \citep{G81a}, and 2) the proper
motions measured there and in W51 Main \citep{G81b}, 
Sgr B2 \citep{RSM88}, W49 N \citep{GMR92}.  Of course 
the \hho\ masers in W3 OH-TW also show a clear bipolar flow 
\citep{AMM93}, and the $\sim50$~\kms\ proper motion of one of the
OH masers from the TW object suggests that at least some of the weak OH 
masers in the new class introduced by \citet{GM87} 
are also associated with strong outflows.

The shock chemistry calculations of \citet{ND89} show that behind 
the front of a dissociative (``$J$'') shock of velocity 80 \kms, OH 
re-forms at a high abundance approaching $[{\rm OH/H}] \approx 10^{-5}$,
while the \hho\ abundance is about an order of magnitude lower.
In contrast, behind a non-dissociative (``$C$'') shock almost all the oxygen
becomes tied up in \hho\ (e.g., Draine, Roberge, \&\ Dalgarno 1983).
The velocities in most \hho-emitting outflows are lower than 50 \kms,
indicating non-dissociative shocks (in whose aftermath conditions
are very conducive for pumping \hho\ masers; \citet{E92}). This
might explain why the \hho\ masers are always much more powerful 
(by up to more than 4 orders of magnitude) than the OH masers in 
W3~OH-TW (and Orion-KL as discussed below).
Alternatively, as mentioned above, the densities necessary
for 22 GHz \hho\ maser action are two orders of magnitude higher
than the values usually invoked for OH masers, and the
OH maser emission should be quenched at these densities.

The masers associated with the TW object and those associated with
source-I in the Orion-KL region (Menten \&\ Reid 1995) have definite similarities.
Both are associated with strong outflows.  They
have weak (or undetectable) thermal continuum emission
at radio wavelengths and no clear, direct IR detections. 
Also both have strong \hho\ and weaker OH masers.
If the OH masers in Orion-KL and W3~OH-TW have a similar
origin, then why is it that toward Orion-KL \citet{JMN89} find 
many dozen features in the 1665 MHz (and about 2 dozen in the 1612 MHz) 
OH line, while we detect only one each in the 1665 and 1667 MHz lines      
(and none in the 1612 MHz line) toward W3~OH-TW?
Given our $(4\sigma)$ detection limit of roughly 0.1~Jy, we would
detect a 2.5 Jy strong feature in Orion put at the roughly
5 times larger distance of W3OH. Thus, we would detect 25 of the Johnston
et al.~1665 MHz features, many of them barely, and one
of their 1612 MHz features. Although no strong correlation between
\hho\ and OH luminosity seems to have been established yet,
we note that the maser luminosity in both species is about an order of 
magnitude greater in Orion-KL than W3 OH-TW, as is the case for the 
infrared luminosity in both regions. This makes the paucity of OH 
features observed toward W3 OH-TW compared to Orion-KL understandable.

\subsection {OH Maser/UCHII Region Association}

In regions of massive star formation, one often finds interstellar OH 
masers.  The traditional view is that interstellar hydroxyl masers
(and class II methanol masers) form in the slowly 
expanding (3 to 10 \kms), dense ($\lax 10^7$~cm$^{-3}$), 
and warm ($\approx150$K) molecular layers driven by the
expanding UCHII region (\eg \citet{EJ78,CW91,M97}), where the OH is likely
formed by photo-dissociation of \hho.  
Since the strongest OH sources are generally associated with UCHII 
regions, most researchers have concluded that OH masers are probably 
associated with massive OB-type (ionizing) stars.   
However, it may well be that the ``traditional'' OH--UCHII region association
picture emerged due to a selection effect, caused by the fact that
earlier searches for OH masers were conducted predominantly 
toward HII regions. 
At least three surveys suggest that this is indeed the case.

In \citet{ARM00}, we selected for observation interstellar OH 
masers whose declinations were above $-45^{\circ}$ and peak flux densities 
were stronger than 1 Jy in both circular polarizations in at least one OH 
main-line transition.  We observed 91 OH masers and noted that our sample was  
at least $80\%$ complete.  Towards each of the observed OH maser regions, we  
also looked for 8.4 GHz continuum emission detectable above a noise level 
of $\approx 0.15$ mJy~beam$^{-1}$.    The Maser--UCHII region association was 
judged by analogy to W3~OH.   In W3~OH, the OH masers closer than $3''$ 
to the center of the brightest 8.4 GHz continuum emission ($\approx 2''$ 
from the edge) appear to be
associated with the UCHII region, while the OH masers farther away 
(i.e., those we have associated with the TW source) are not.   At a 
distance of 2.2~kpc (Humphreys 1978) 
to W3~OH, $2''$ corresponds to 0.02 pc.   We then examined all sources
in the \citet{ARM00} survey and found that in only 49 of the 91 maser fields 
($54\%$) were {\it any} of the OH masers closer than 0.02 pc to the edge of an 
UCHII region. 

In another recent survey, \citet{FC00} presented a search for 8.5 GHz radio 
continuum emission toward 45 sites of OH and \hho\ maser emission with 
Galactic longitudes in the range $352\degg52$ to $24\degg79$.  Only toward
17 of them did they find continuum emission above a detection threshold of 
approximately 0.5 mJy~beam$^{-1}$.  (At this frequency, according to 
\citet{FC00}, an ionization-bounded, optically thin, HII region produced by a 
ZAMS star of spectral type B1 at a distance of 5 kpc is 1 mJy and would have 
been detected.) 
If one considers only the OH maser regions and uses the maser-continuum 
association criterion described above, one finds that only 16 of the 26 
($62\%$) regions that contain OH masers  have associated continuum emission.  

Finally, \citet{GM87} studied the ground state OH maser and 15 GHz 
continuum emission associated with 11 regions of star formation, demanding only
that the 1720 MHz OH transition be present when selecting regions for study.   
They noted that $75-80\%$ of the maser clusters within their 11 regions 
appeared to be closely associated with HII emission.  Detection limits 
(rms noise levels) varied widely from source to source but typically were in
the range of a 0.3 mJy~beam$^{-1}$ to 3  mJy~beam$^{-1}$.  
Although this percentage is not directly comparable to the percentages 
from the two previous surveys, since we did not attempt to subdivide maser 
regions into maser clumps when discussing those surveys, it does strengthen 
the argument that many OH masers have no associated continuum.  
Gaume \& Mutel (1987) grouped these unassociated masers into a separate 
``class'' and noted that they tended to be much weaker than other OH masers.  

How does one explain the presence of OH masers with no associated continuum?  
At least two possibilities come to mind.  First, the masers may be associated 
with a very young HII region that is currently below the detection limit
because of its small size.  While one might expect this phase in the
development of an UCHII region to be very brief, \citet{Keto2002}
show that the strong gravitational attraction of material very close to
a massive star can significantly extend the lifetime of this phase.  
In the future, these HII regions can be expected to grow larger and become 
detectable.  An alternative possibility for the absence of an UCHII region 
near some OH masers is that these masers may be associated with a star that 
will never give rise to an HII region.  The TW source and other outflow 
regions, for example, may contain a central star less massive than a 
B3 star and not produce sufficient ionizing photons to create a detectable
UCHII region.

It is uncertain how many of the OH masers with no detected continuum emission 
are similar to the TW object masers.  In general, OH masers associated 
with outflows can be expected to have weak features spanning tens of 
\kms\ in radial velocity.  In the survey of \citet{ARM00}
$\approx 20$\% of interstellar OH maser spectra have velocity spreads 
$\gax10$~\kms\ and $\approx 5$\% have velocity spreads $\approx30$~\kms, 
the latter being inconsistent with the expansion velocity of a compact 
HII region.  Unfortunately, it is possible that 
most surveys could have missed weak, high velocity, maser features
owing to bandwidth and sensitivity limitations.  

While the spread in radial velocity of the OH masers toward W3~OH is 
not large ($\approx9$~\kms), the observed proper motion of the 1665 MHz maser 
toward the TW source is very large, implying that the outflow in this 
source lies close to the plane of the sky.  This is corroborated by the 
proper motions measured for the TW \hho\ masers by \citet{AMM93}.  
This may not be a coincidence, since there may be a preferential 
orientation for observing detectable masers in a bipolar outflow.
For outflows, the longest gain paths are likely to be perpendicular 
to the outflow axis.   Thus, these masers may be strong only when the 
outflow axis is nearly perpendicular to our line of sight, and they
would show small Doppler shifts from the systemic velocity.
If this effect is important, then a significant fraction of the OH 
masers with no detected continuum emission could be associated with outflows.


\begin{figure}
\epsscale{0.89}
\plotone{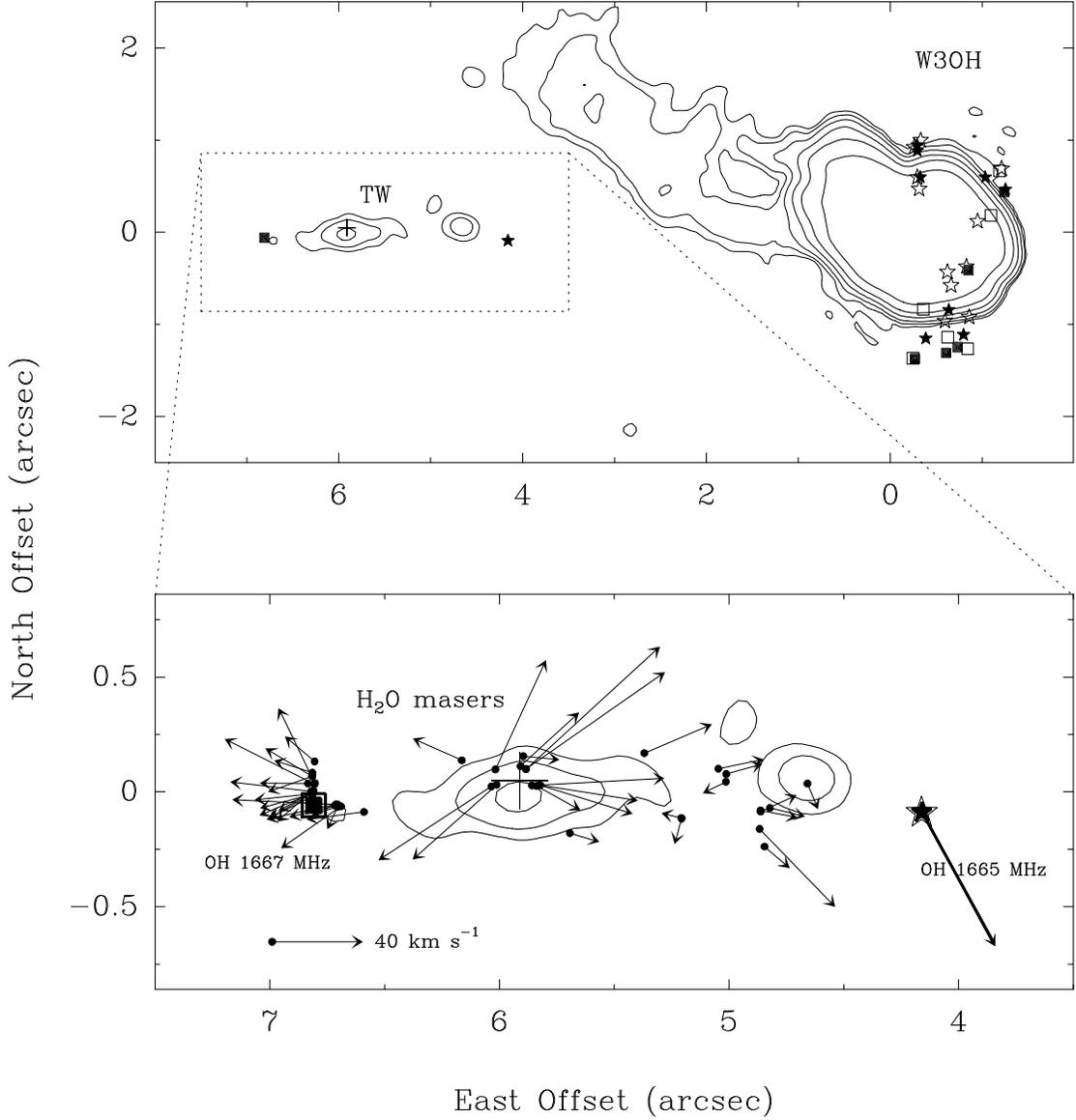}
\figcaption[]{ 
{\it (Top panel)} 1665 and 1667 MHz OH masers from \citet{ARM00}
and two ``outlying'' OH masers (this paper) superimposed on the 8.4 GHz VLA 
contour plot of Wilner et al. (1999).  The ``stars'' represent 1665 MHz OH 
maser emission and the ``squares'' 1667 MHz OH maser emission.  Open symbols 
are RCP and closed symbols LCP.  `TW' stands for the Turner-Welch object 
(Turner \& Welch 1984). {\it (Lower panel)} H$_2$O and outlying OH maser 
emission near the Turner Welch object.  Positions and proper motions of the 
H$_2$O masers are indicated by filled circles and light arrows, 
respectively (Alcolea {\it 
et al.} 1993) and the large cross marks the \hho\ maser center of 
expansion determined by these authors. Positions of the OH masers are
indicated by either a ``star'' or a 
``square'' (this paper) and the single proper motion by a heavy 
arrow. The origin of coordinates corresponds to 
$\alpha_{B1950} = 02^h 23^m 16\s499, \delta_{B1950} = 61^{\circ} 38' 57\as21$. 
\label{fig2} }
\end{figure}

\section{Conclusions}
During an interferometric survey of OH masers in star-forming regions, we 
detected weak OH maser emission associated with the Turner-Welch object,
a very deeply embedded protostellar object. 
Using archival VLBI data, we were able to determine the proper motion of 
one of the maser features, which is of comparable magnitude and 
direction as the proper motions of the \hho\ masers,
which arise from a bipolar outflow centered on the TW object.
While conventional wisdom associates OH masers with the slowly expanding 
molecular envelopes of ultracompact HII regions, we argue that the W3 OH/TW OH
masers are part of a new class of OH masers that trace protostellar outflows.
Another example of this class of OH masers might be the 
Orion-KL/IRc~2/Source-I masers and, possibly, some of the sources mapped 
by \citet{FC00} in which OH and \hho\ masers are coexistent and are 
offset from continuum emission.
 
The near spatial coincidence of OH and \hho\ maser emission in outflow
sources may be explained by OH production by photo-dissociation of \hho\
by the interstellar radiation field.
The densities required for \hho\
maser action are roughly two orders of magnitude higher than those suitable 
for OH masers.  This is consistent with the observation that the OH maser
spots are located at the outer edges of the \hho\ emission region.
One important avenue of future research suggested by our results
is high spatial resolution observations of this class of OH masers, 
since OH is an important chemical constituent of post-shock gas not 
otherwise observable with comparable resolution.

\end{document}